\documentclass[12pt,preprint]{aastex}
\usepackage{times}

\newcommand{\twe}{$\tau_\mathrm{we}$}
\newcommand{\tns}{$\tau_\mathrm{ns}$}

\shorttitle{Center-to-Limb Variation in Helioseismic Travel Times}
\shortauthors{Zhao et al.}

\begin{document}
\title{Systematic Center-to-Limb Variation in Measured Helioseismic Travel 
Times and Its Effect on Inferences of Solar Interior Meridional Flows}

\author{Junwei Zhao\altaffilmark{1}, Kaori Nagashima\altaffilmark{1}, 
R.~S.~Bogart\altaffilmark{1}, A.~G.~Kosovichev\altaffilmark{1}, and 
T.~L.~Duvall, Jr.\altaffilmark{2}}
\altaffiltext{1}{W.~W.~Hansen Experimental Physics Laboratory, Stanford 
University, Stanford, CA94305-4085}
\altaffiltext{2}{Solar Physics Laboratory, NASA Goddard Space Flight 
Center, Greenbelt, MD20771}

\begin{abstract}
We report on a systematic center-to-limb variation in measured 
helioseismic travel times, which must be taken into account for an 
accurate determination of solar interior meridional flows. The systematic 
variation, found in time-distance helioseismology analysis using 
{\it SDO}/HMI and {\it SDO}/AIA observations, is different in both 
travel-time magnitude and variation trend for different observables. 
It is not clear what causes this systematic effect.
Subtracting the longitude-dependent east-west travel times, obtained 
along the equatorial area, from the latitude-dependent north-south 
travel times, obtained along the central meridian area, gives remarkably 
similar results for different observables. We suggest this as 
an effective procedure for removing the systematic center-to-limb variation. 
The subsurface meridional 
flows obtained from inversion of the corrected travel times are approximately
10~m~s$^{-1}$ slower than those obtained without removing the systematic
effect. The detected center-to-limb variation may have important 
implications in the derivation of meridional flows in the deep interior,
and needs a better understanding.
\end{abstract}


\section{Introduction} 
An accurate determination of the meridional flow speed in both the solar
photosphere and the solar interior is crucial to the understanding of 
solar dynamo and predicting solar cycle variations. For example, 
\citet{dik06} suggested that the slow-down of the meridional circulation 
during the solar-cycle maximum could change the duration of the following 
minimum and delay the onset of the next solar cycle. \citet{hat10} found 
that the meridional flow speed was substantially faster during the solar 
minimum of Cycle 23 than during the previous minimum, and suggested that 
this might explain the prolonged minimum of Cycle 23. Using flux-transport 
dynamo model simulations, \citet{dik10} suggested that the prolonged 
minimum of Cycle 23 might be due to that the poleward meridional flow extended
all the way to the pole in Cycle 23, unlike in Cycle 22 the flow switched to
equator-ward near the latitude of $60\degr$. All these works demonstrate
that an accurate measurement of the meridional flow speed is very important.
The photospheric 
meridional flow speed can be determined by tracking certain 
photospheric features, such as magnetic structures and supergranules 
\citep[e.g.,][]{kom93, hat10, hat10b, giz04, sva06}, although it is 
not quite clear how well the motions of these surface features represent 
the photospheric plasma flows. The photospheric meridional flow speed can
also be inferred from Doppler-shift measurements \citep[e.g.,][]{hat96,
ulr10}, and recently \citet{ulr10} made extensive comparisons of 
the meridional flow speeds obtained by different methods. The meridional 
flows in the solar interior are primarily determined by helioseismology,
i.e., by measuring frequency shifts between poleward and equator-ward
traveling acoustic waves \citep[e.g.,][]{bra98, kri07, rot08}, and by use of 
local helioseismology techniques, namely, ring-diagram analysis and 
time-distance helioseismology \citep[e.g.,][]{gil97, cho01, hab02, 
zha04, gon08}.

The subsurface meridional flow speeds obtained by the two different 
local helioseismology techniques are in reasonable agreement for at least 
the upper 20 Mm of the convection zone \citep[e.g.,][]{hin04}. However, 
the agreement between the two analysis techniques cannot rule out 
that both techniques may be affected by the same or similar systematic 
effects. In this Letter, we report on a systematic center-to-limb 
variation in helioseismic travel times measured by the time-distance 
helioseismology technique, which was previously unnoticed but must 
be taken into account in the inference of the subsurface meridional flows. 
Other helioseismology techniques, such as the ring-diagram analysis, may 
also be affected by a similar systematic effect (Rick Bogart, private communication). 
We develop an empirical correction procedure by measuring the center-to-limb 
variation along the equatorial area during the periods when the solar 
rotation axis is perpendicular to the line-of-sight, i.e., when the solar 
B-angle is close to $0\degr$. This correction scheme provides consistent 
results for the acoustic travel times in the north-south directions measured 
from different HMI observables and AIA chromospheric intensity variations. 
We introduce our data analysis procedure and present results in \S2, and 
discuss the results and their implications in \S3. 

\section{Data Analysis and Results}
\label{sec2}

\subsection{Data Analysis Tools} 
\label{sec21}
To facilitate analysis of the large amount of data from the 
Helioseismic and Magnetic Imager \citep[HMI;][]{scherrer12, schou12} 
onboard {\it Solar Dynamics Observatory} \citep[SDO;][]{pes12}, a 
time-distance helioseismology data-analysis pipeline was developed 
and implemented at the HMI-AIA Joint Science Operation Center 
\citep{zha12}. Every 8 hours, the pipeline provides measurements of 
acoustic travel times, and generates maps of subsurface flow and 
wave-speed perturbations by inversion of the measured travel times, 
covering nearly the full-disk Sun with an area of $120\degr\times120\degr$. 

For the pipeline processing, we select 25 overlapping areas on the solar disk 
and analyze the 8-hr sequences of solar oscillation data separately for each
region. Acoustic travel-time maps are obtained for 11 selected wave travel 
distances, and are inverted to derive maps of subsurface velocity and 
wave-speed perturbations from the surface to about 20~Mm in depth 
\citep{zha12}. Then the 
results for individual regions are merged into nearly-full-disk maps covering 
$120\degr$ in both longitude and latitude, with a spatial sampling of 
$0\fdg12$ pixel$^{-1}$ on a uniform longitude-latitude grid. The pipeline 
gives acoustic travel time measurements from two different fitting techniques
\citep{cou12}, and inversion results based on ray-path and Born-approximation 
sensitivity kernels. Both measurement uncertainties for different 
distances and inversion error estimates for different inversion depths
are given in \citet{zha12}. In this Letter, 
only the acoustic travel times obtained from Gabor-wavelet fitting 
\citep{kos97} and inversion results based on the ray-path approximation 
kernels are presented.

Although the pipeline is designed to analyze the HMI Dopplergrams,
it can nevertheless be used to analyze other HMI observables 
that carry solar oscillation signals, e.g., continuum intensity, line-core 
intensity, and line-depth. The full-disk data from the 1600~\AA\ and 
1700~\AA\ channels of Atmospheric Imaging Assembly \citep[AIA;][]{lem12} 
onboard {\it SDO} can also be used for helioseismology studies with 
good accuracy \citep{how11}. In this study, we compare 
results obtained from the four HMI observables and the AIA 1600~\AA\ 
data following the same analysis procedure. This comparison helps us to 
identify the systematic center-to-limb variation and develop a correction 
method.

\subsection{Center-to-Limb Variation in Measured Travel Times}
\label{sec22}
The north-south (\tns) and west-east (\twe) acoustic travel-time differences
approximately represent the north-south and west-east flow components, 
respectively, although a full inversion is required to determine more 
precisely these flows. We first show the measured travel 
time differences and then present the inversion results. We choose a 10-day 
period of December 1 through December 10, 2010, when the solar B-angle 
between the equator and the ecliptic is close to $0\degr$ to avoid 
complications caused by leakage of the solar rotation signal into 
the meridional flow measurements.

\begin{figure}
\epsscale{0.95}
\plotone{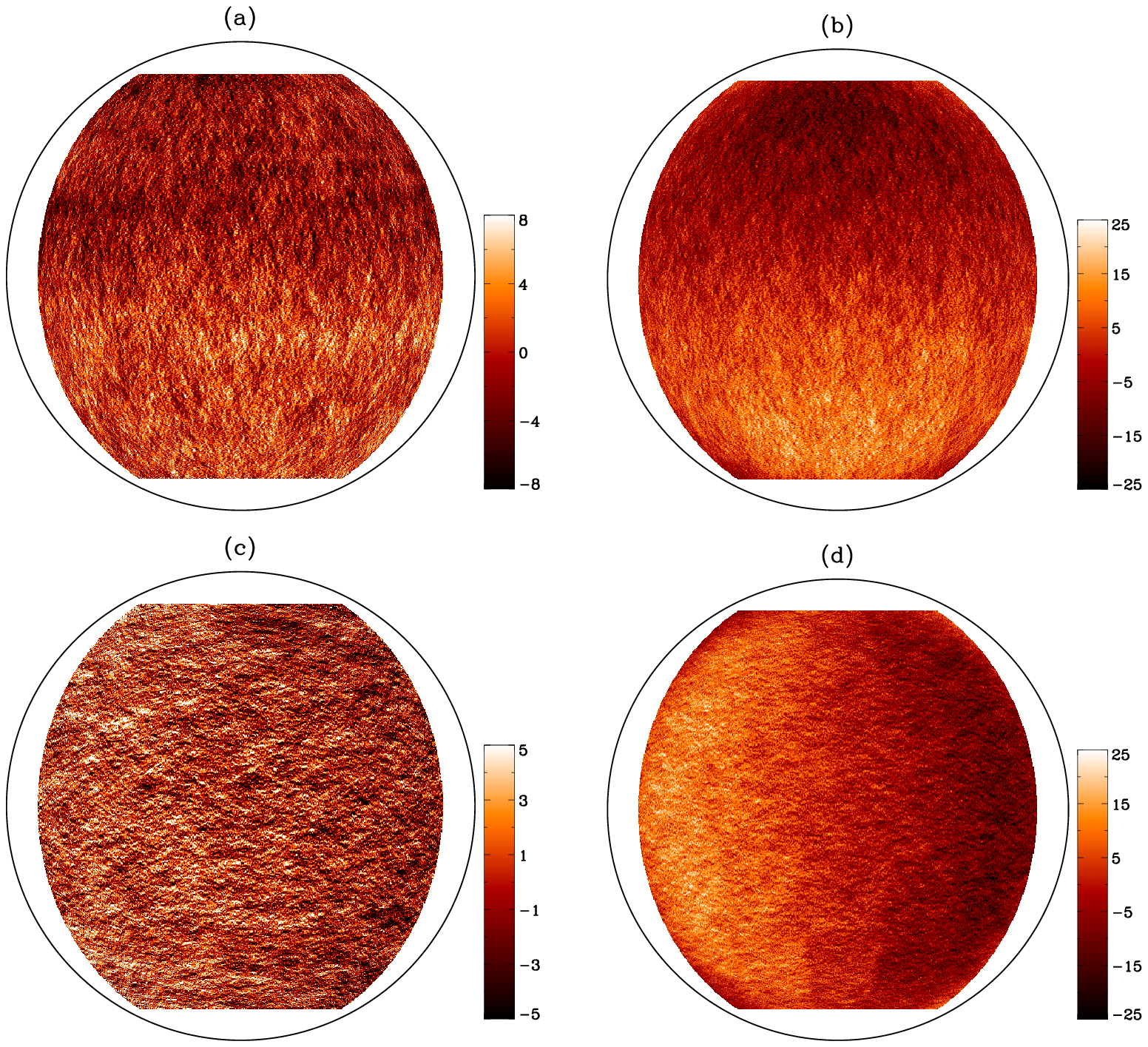}
\caption{Travel-time maps (in unit of sec), covering $120\degr$ in both 
longitude and latitude, for \tns\ obtained from (a) HMI Dopplergrams and 
from (b) HMI continuum intensity, and for \twe\ obtained from (c) Dopplergrams 
and from (d) continuum intensity after a longitudinally averaged profile,
representing the differential rotation, is removed through all longitudes. 
Note that color scales are not the same in all panels. }
\label{map}
\end{figure}

The upper panels of Figure~\ref{map} show the nearly-full-disk map of \tns,
averaged over the 10-day period, measured from the HMI Dopplergram and 
continuum intensity data for an acoustic travel distance of $1\fdg08 - 1\fdg38$.
The general pattern of positive \tns\ in the southern hemisphere
and negative \tns\ in the northern hemisphere is usually thought
to be caused by the interior poleward meridional flows. However, the
apparent differences between the magnitude of \tns\ obtained from the Doppler
data and that obtained from the continuum intensity data indicate 
that there are additional systematic variations. The lower panels 
show the averaged \twe\ maps measured from the same observables 
after the latitude-dependent travel times caused by the differential 
rotation, obtained by averaging the measurements of all longitudes, are 
subtracted.
One would expect \twe\ be relatively flat along same latitudes because 
the solar rotation does not vary significantly with longitude, but both 
\twe\ maps show systematic travel-time variations along the same latitudes, 
positive in the eastern hemisphere and negative in the western hemisphere. 
The longitudinal variation 
is quite significant for the measurements from the intensity data 
(Figure~\ref{map}d) and small but not negligible for the measurements 
from the Dopplergram data (Figure~\ref{map}c). 

To more quantitatively illustrate this systematic variation, we average 
\tns\ in a $20\degr$-wide band along the central meridian as a function of 
latitude, and display the averaged curves in the top panels of 
Figure~\ref{curves} for three selected measurement distances and
for the four HMI observables: Doppler velocity, continuum intensity,
line-core intensity, and line-depth. In the middle-row panels, we show 
the corresponding \twe\ curves obtained by averaging over a $20\degr$-wide 
band along the equator as a function of longitude (hereafter, longitude 
is relative to the central meridian). If there were no systematic 
center-to-limb variation in the measured acoustic travel times, 
\tns\ obtained from the different observables would agree with each 
other, and \twe\ would remain flat for all observables. But clearly, 
the measurements are not as expected. Among all these measurements, the 
\tns\ curves 
show not only different magnitude of travel-time shifts but sometimes 
also opposite 
variation trends. For \twe, it is found that the measurements from the
Dopplergrams have the smallest systematic variations and are similar 
to the line-core intensity measurements at larger distances. However, 
the travel times measured from the continuum intensity and line-depth data show 
not just substantial travel-time shifts for all measurement distances but 
also opposite center-to-limb variations. It is quite clear that the travel-time 
variations along the equator are not caused by solar flow but represent
a systematic center-to-limb variation. If we treat the longitude-dependent 
\twe\ variations along the equatorial area as the systematic center-to-limb 
variations, and subtract these variations from the latitude-dependent \tns\
measured using the corresponding observables, 
we get the residual curves shown in the bottom-row panels of 
Figure~\ref{curves}. It is remarkable that for all the measurement 
distances, the results from all four HMI observables are in reasonable 
agreement. This suggests that the residual travel times correspond
to the subsurface meridional flow signals.

It was demonstrated that the {\it SDO}/AIA 1600~\AA\ data are also suitable
to perform helioseismology analysis \citep{how11}. Figure~\ref{AIA} 
shows a comparison of the results obtained from the HMI Dopplergrams and 
the AIA 1600~\AA\ data for one selected measurement distance, $3\fdg12 -
3\fdg84$. Similarly, 
\tns\ measured from the AIA 1600~\AA\ data also differ from that measured 
from the HMI Dopplergram. The \twe\ measured from the AIA data also show 
systematic variations, though its longitudinal trend is opposite to that
measured from HMI continuum intensity and the variation magnitude is 
much smaller. The residual, obtained after subtracting the 
longitude-dependent \twe\ from the latitude-dependent \tns\ for AIA 
1600~\AA, is quite similar to that obtained from the HMI Dopplergrams
and other HMI observables. This result is particularly remarkable because 
HMI and AIA are different instruments observing in different spectral 
lines: AIA in 1600~\AA\ and HMI in \ion{Fe}{1}~6173~\AA, which are formed 
at different heights in the solar atmosphere.

\subsection{Effect on Meridional Flow Inversions}
\label{sec23}
The systematic center-to-limb travel-time variations would have 
an apparent effect on the inference of subsurface meridional flows 
obtained by inversion. Here, we investigate how the inferred
meridional flow speed changes after removal of this systematic effect.

We employ the time-distance helioseismology pipeline code developed for 
inversion of the acoustic travel times \citep{zha12}. 
The inversion results are shown in Figure~\ref{invert}. The middle row
of Figure~\ref{invert} presents inversion results obtained from travel-time
measurements shown in Figure~\ref{curves}, but with an extension of 
$15\degr$ closer to both poles and both limbs to better illustrate that 
the removal of the systematic effect also helps to improve inversions 
in high-latitude areas. The top and bottom rows of Figure~\ref{invert}
show results obtained from June 1 -- 10, 2010 and June 1 -- 10, 2011,
when the solar B-angle were also close to 0$\degr$. It can be seen 
from left columns of Figure~\ref{invert} that above $\sim 55\degr$ 
latitude, the inferred meridional flow drops in speed and becomes 
equator-ward for the depth of 0 -- 1 Mm, but remains pole-ward deeper
than 3 Mm. This is a suspicious behavior for the meridional flows. 
We then antisymmetrize the east-west direction velocity caused by 
the center-to-limb variations (middle column of Figure~\ref{invert}), 
and subtract it from the inferred meridional velocity. The residual 
subsurface meridional flows show much more consistent behaviors at different
depths in high-latitude areas (right column of Figure~\ref{invert}).
From Figure~\ref{invert}, one can also find that the center-to-limb 
variation measured from equatorial area also slightly change with 
time, and why there is such a change is not understood and worth further
studies.

The removal of the center-to-limb variation in the measured travel times 
also decreases the inferred flow speed by nearly 10~m~s$^{-1}$.
This indicates that the subsurface meridional flows derived 
from the previous time-distance studies \citep[e.g.,][]{zha04} might 
have overestimated the flow speed by a large fraction. 
Figure~\ref{all_merid} shows that after removal of the systematic effect,
the meridional flow speed is closer in magnitude to the results obtained
by the magnetic feature tracking \citep{hat10} and surface Doppler 
measurements \citep{ulr10}. Note that the results from the magnetic 
feature tracking and from the MWO Doppler observations shown in 
Figure~\ref{all_merid} are both averaged over a 6-month period with 
our analyzed 10-day period in the middle. The difference of analysis 
period may result in some differences seen in the figure. The meridional
flows obtained from this study are displayed in Figure~\ref{all_merid}
after a $2\degr$ spatial averaging to remove the strong fluctuations 
caused by supergranular flows.

\section{Discussion}
\label{sec3}
The analysis of acoustic travel times obtained from different observables 
of the HMI and AIA instruments on SDO by the time-distance helioseismology
technique has revealed a systematic center-to-limb variation.
The systematic variation
is different for different observables, and range from $\sim$2~sec for
the HMI Dopplergram to $\sim$10~sec for the continuum intensity measurements. 
For an accurate determination of subsurface meridional flows, and also for 
more accurate inference of full-disk subsurface flow fields, 
this systematic effect should be removed. We have developed an empirical
correction procedure by removing the systematic variation measured along 
the equatorial area during the period when solar B-angle is close to $0\degr$. 
This correction reconciles the latitude-dependent \tns\ measured from 
different observables and reduces the inferred meridional flow speed by
about 10~m~s$^{-1}$.

It is not quite clear what causes this systematic center-to-limb
effect. However, this effect is unlikely caused by instrumental or 
data calibration, because it is observed in the data from two different 
instruments, HMI and AIA, and it exists in different observables of 
HMI. We also rule out the following factors as causes of this effect: 
finite speed of light, contribution of horizontal wave component, 
and foreshortening effect. Due to the finite speed of light, acoustic 
wave signals observed at high latitude and observed near the equator 
are not simultaneous, but this effect is negligible given 
the small measurement distances used in this study. The horizontal wave 
component \citep{nig07} is used to explain the center-to-limb variation 
in mean travel times  observed by \citet{duv03}, but this effect is 
not significant in travel-time differences for waves traveling in 
opposite directions and in small distances as used in this study. 
To check foreshortening effects, we mimicked the high-latitude data 
by reducing the spatial resolution of the data observed at lower latitudes, 
and our measurements did not show systematic changes in the measured 
travel times similar to the center-to-limb variations reported here.

It is more likely that this effect has a physical origin related to 
properties of solar acoustic waves in the solar atmosphere or response 
of the spectral lines to the solar oscillations. It is well known that 
for a given spectral line the Sun is observed at the same optical depth 
but not at the same geometrical height. The observed height gradually 
increases with distance from the disk center. Thus it is possible that 
the acoustic waves traveling in opposite directions and different 
height will give different measured travel times due to the subsurface 
location of the wave source and the waves evanescent behavior above
the photosphere \citep{nag09}. However, this cannot explain why different 
observables give different magnitude and trend of the center-to-limb 
variations. Another interesting fact is that the largest systematic 
effect exists in the measurements using HMI continuum that forms at the lowest
height in the atmosphere. As one moves up to the height where Doppler 
velocity is measured, the systematic effect is reduced. As the measurements 
move further up to where AIA~1600~\AA\ is formed, the effect reverses sign.
This may give us some indication of that this center-to-limb variation
is related to the line formation height. It is also possible that 
this systematic effect is related to differences in the acoustic power 
distributions, line-asymmetry of solar modes \citep{duv93}, and the correlated 
noise effect \citep{nig98}. The cause of this systematic effect is worth 
further studies, and may be resolved by numerical modeling of solar 
oscillations including the spectral line-formation simulation and 
spherical geometry of the Sun. It is worth attention that when inferring
meridional flow velocity from surface Doppler measurements, a systematic
center-to-limb velocity profile also needs to be removed \citep{ulr88,
ulr10}. A similar limb-shift effect very likely exists in the HMI 
Dopplergrams that are used in our time-distance measurements, and
it is not clear whether this effect accounts for some of the systematic 
variations in our measured acoustic travel times. This is worth further 
studies.

\begin{figure}
\epsscale{0.95}
\plotone{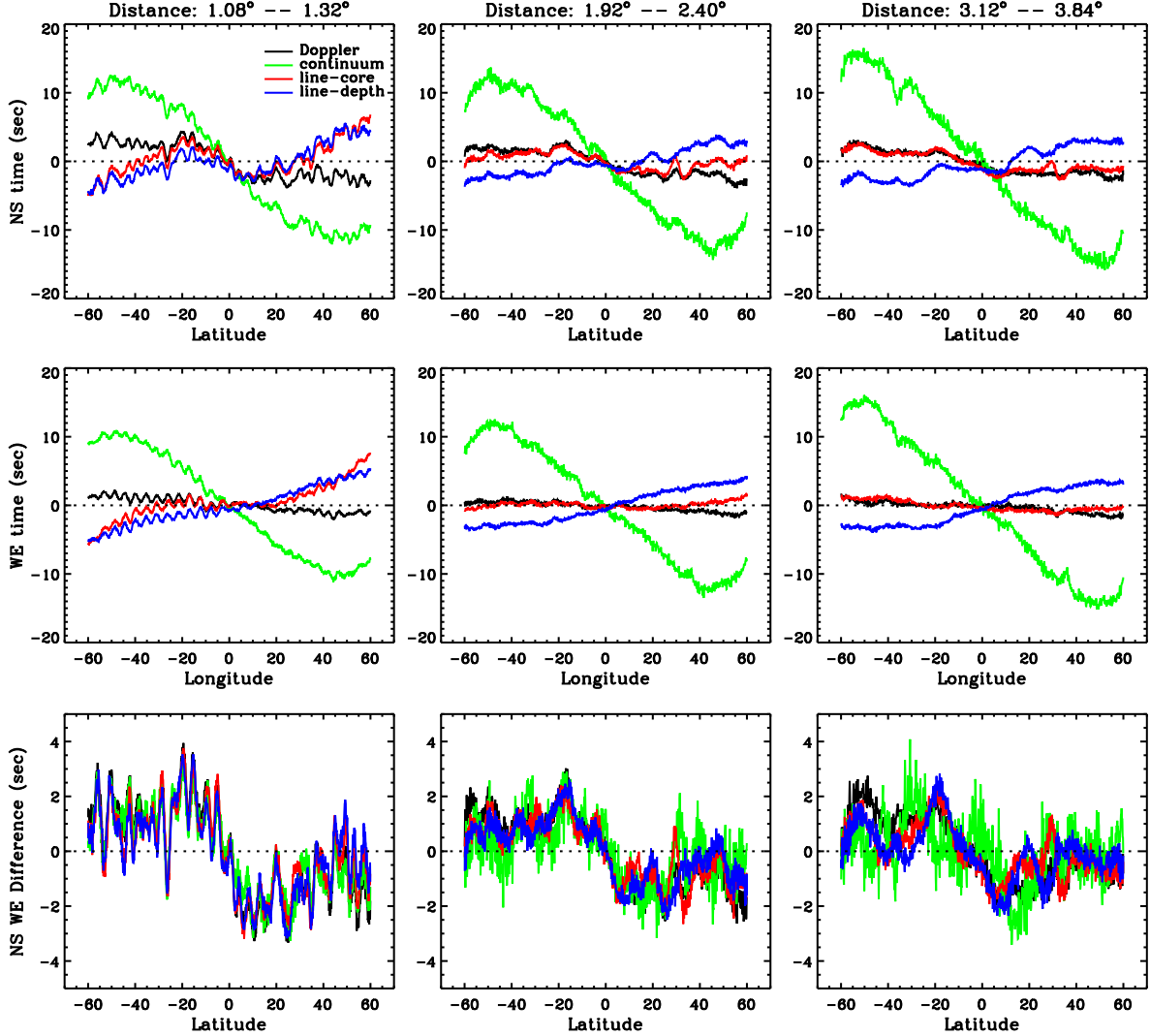}
\caption{{\it Top:} Averaged curves of latitude-dependent \tns, obtained 
from different HMI observables and for different measurement distances. 
{\it Middle:} Averaged curves of longitude-dependent \twe, obtained 
from different observables and for different measurement distances. 
{\it Bottom:} Differences of \tns\ and \twe. Note that the 
vertical scales for the upper two rows are different from those for the 
bottom row.}
\label{curves}
\end{figure}

\begin{figure}
\epsscale{0.95}
\plotone{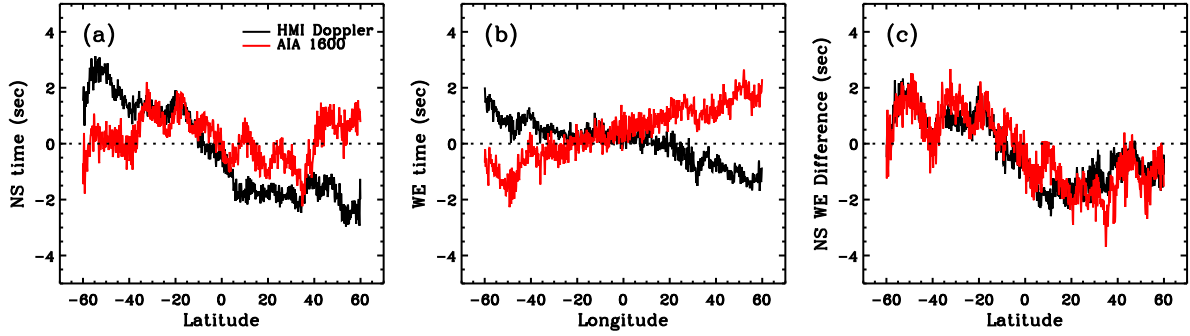}
\caption{(a) Averaged latitude-dependent \tns, obtained from HMI Dopplergrams 
and from AIA 1600~\AA\ data for a measurement distance of $3\fdg12 - 3\fdg84$. 
(b) Averaged longitude-dependent \twe\ for these two observables. (c) 
Differences of \tns\ and \twe. }
\label{AIA}
\end{figure}

\begin{figure}
\epsscale{0.95}
\plotone{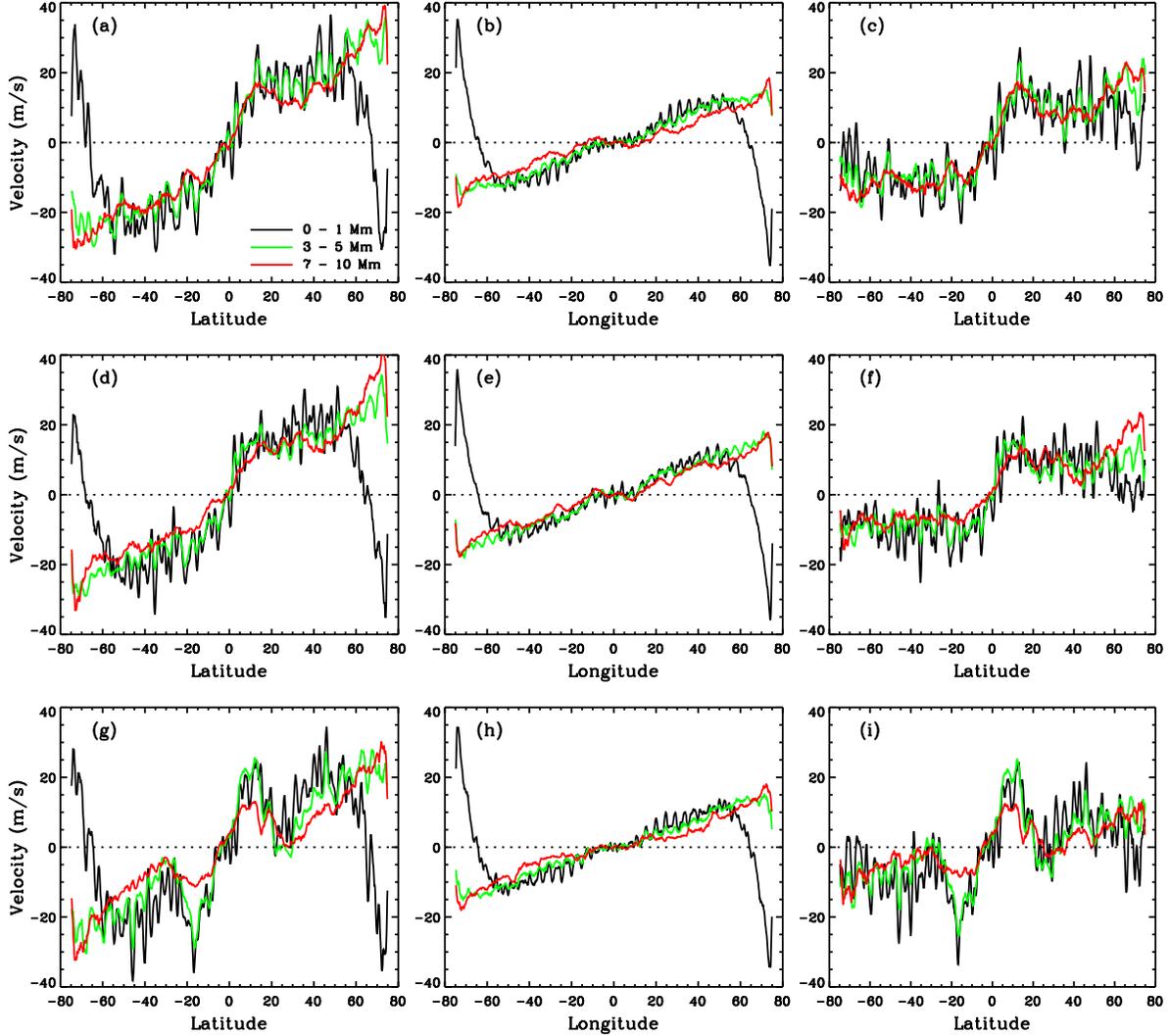}
\caption{Results obtained from the periods of June 1 -- 10, 2010 (top row),
December 1 -- 10, 2010 (middle row), and June 1 -- 10, 2011 (bottom row).
(a) Meridional flow velocity for different depths, obtained by
inversions of acoustic travel times measured from the HMI Dopplergram 
and without removal of the systematic center-to-limb variations. (b) 
Antisymmetrized east-west velocity obtained by inversion, representing
the center-to-limb variations. (c) Meridional flow velocities after 
removal of the systematic center-to-limb variations. (d) -- (f) and 
(g) -- (i) are the same as (a) -- (c) but for different time periods. }
\label{invert}
\end{figure}

Although the cause of this systematic center-to-limb variation is 
not well understood, it is demonstrated that the empirical correction 
procedure can improve the inferred subsurface meridional flows.
Figures~\ref{curves} -- \ref{invert} demonstrate 
that subtracting the longitude-dependent \twe\ from the latitude-dependent 
\tns, measured from same observables, is an effective way to remove 
this systematic effect. This correction procedure helps to reconcile 
the \tns\ measured from different observables, and also helps to remove 
the inconsistent behaviors of meridional flows in high-latitude 
areas. As an effect, the newly obtained subsurface 
meridional flows at shallow depths are approximately 10~m~s$^{-1}$
slower than the speed previously derived following
a similar analysis procedure.  
This systematic center-to-limb variation in measured acoustic travel times
has an important implication in the long-searched deep equator-ward
meridional flows, and this is currently under investigation.

\begin{figure}
\epsscale{0.7}
\plotone{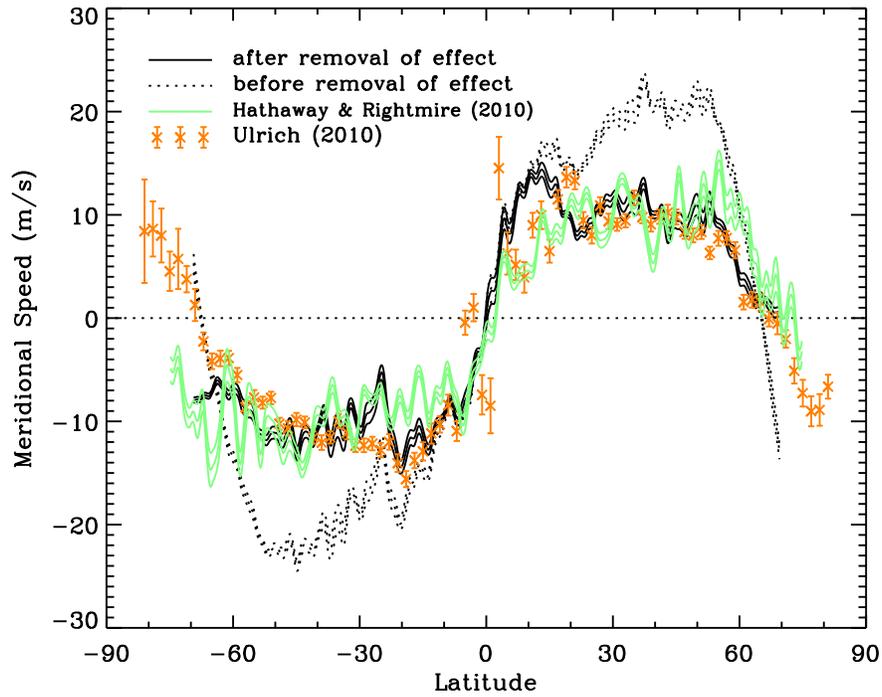}
\caption{Comparison of the meridional flow speed obtained by magnetic feature 
tracking \citep{hat10}, from surface Doppler measurements \citep{ulr10}, 
and obtained in this study for a depth of 0 -- 1 Mm before and after 
the removal of the systematic effect. Our results are from a 10-day
period, December 1 -- 10, 2010, and are displayed after spatial averaging.
The results from magnetic feature tracking and from Doppler measurement 
are averaged over a 6-month period with our analyzed period in the middle.}
\label{all_merid}
\end{figure}

\acknowledgments
SDO is a NASA mission, and HMI project is supported by NASA contract 
NAS5-02139. We thank Drs.~Roger Ulrich and David Hathaway for providing 
us their analysis results used to make Figure~\ref{all_merid}. We also
thank an anonymous referee whose suggestions help to improve the quality
of this paper. 


\newpage

\newpage

\newpage

\newpage

\newpage

\end{document}